\documentclass{jaa}
\usepackage{natbib}
\usepackage{color}
\usepackage{graphicx}

\newcommand{\ty}[1]{\mbox {TYC~8351-1081-1}}
\newcommand{\as}[1]{\mbox {ASAS~J210406-0522.3}}
\newcommand{\tyc}[1]{\mbox {TYC~8351-1377-1}}

\begin{document}\sloppy

%%paper title
%%For line breaks\\ can be used within title
\title{Simplified Method for the Identification of Low Mass Ratio Contact Binary Systems that are Potential Red Nova Progenitors}

%%author names are separated by comma (,)
%%use \and before the last author name
%%use a * along with the number separated by comma
%% for the  author for correspondence
%%\textsuperscript{number} is used for affiliation
%%\affilOne, \affilTwo etc., upto \affilTwentyfive is possible
%%Please note the first letter after \affil is capitalised in the command
%%

\author{Surjit S. Wadhwa\textsuperscript{1*}, Ain Y. De Horta\textsuperscript{1}, Miroslav D. Filipovi\'c\textsuperscript{1}, Nick F. H.  Tothill\textsuperscript{1}, Bojan Arbutina\textsuperscript{2}, Jelena Petrovi\'c\textsuperscript{3} and Gojko Djura\v sevi\'c\textsuperscript{3}}
\affilOne{\textsuperscript{1}School of Science, Western Sydney University, Locked Bag 1797, Penrith, NSW 2751, Australia.\\}
\affilTwo{\textsuperscript{2}Department of Astronomy, Faculty of Mathematics, University of Belgrade, Studentski trg 16, 11000 Belgrade, Serbia.\\}
\affilThree{\textsuperscript{3}Astronomical Observatory, Volgina 7, 11060 Belgrade, Serbia\\}
%\affilTwo{\textsuperscript{2}Department of Q, University Z, Place Pincode, Country.}

%%escape two column mode for title, affiliation and abstract
%%by giving \twocolumn command as shown

\twocolumn[{

\maketitle

%%include \corres to print the corresponding author Email id
\corres{19899347@student.westernsydney.edu.au}

%%include \msinfo for
%%manuscript information such as
%%received, revised and accepted dates
%%
%\msinfo{1 January 2015}{1 January 2015}

%%abstract
\begin{abstract}
The study presents a simplified method to identify potential bright red nova progenitors based on the amplitude of the light curve and infrared (J-H) colour of a contact binary system. We employ published criteria for contact binary orbital instability to show that the amplitude of the light curve for a given contact system with a low mass ($< 1.4M_{\odot}$) primary must be less than a specified value for it to be potentially unstable. Using this we search the photometric data of a large survey to identify about 50 potential bright red nova progenitors. We analyse the survey photometry of each to determine the mass ratio and from the estimated mass of the primary other physical parameters of the systems. We show that each system has physical characteristics indicating potential orbital instability. Using the absolute parameters from our sample we model the expected instability separation and period for low mass contact binary systems.
\end{abstract}

%%insert keywords separated by 3 hyphens using \keywords{words}

}]
%%close the twocolumn escape here

%%include \doinum{number}for the DOI number in the header
%%include \volnum{number} for the volume number in the header
%%include \year{yyyy} for  year of publication in the header
%%include \pgrange{num--num} page range of article in the header
%%include \artcitid{num} for the article citation id
%%include \lp to print last page of the article
%%include \setcounter{page}{pagenum} for the exact starting page of the article

\doinum{12.3456/s78910-011-012-3}
\artcitid{\#\#\#\#}
\volnum{000}
\year{0000}
\pgrange{1--}
\setcounter{page}{1}
\lp{1}

\section{Introduction}
The number of known contact binary systems has, and still is, growing at a phenomenal rate given the new discoveries resulting from sky surveys. As an example, All Sky Automated Survey (ASAS) \citep{2002AcA....52..397P} and the All Sky Automated Survey - Super Nova (ASAS-SN) \citep{2014ApJ...788...48S, 2020MNRAS.491...13J} have added more than 100000 new discoveries. Theoretical models such as  \cite{2003ApJ...582L.105S, 2012JASS...29..145E, 2011A&A...531A..18S, 1977MNRAS.179..359R} predict that contact binary systems with extremely low mass ratios are likely to merge into a single rapidly rotating relatively cool giant star. The merger event is thought to result in a transient nova like event that evolves to remain bright in the red and infrared bands. The event is usually termed a red nova. Although Galactic merger events are predicted to occur commonly (once every 2-3 year), brighter events that are likely to be available for study are more limited at once a decade \citep{2014MNRAS.443.1319K}. There has been only one confirmed observation linking typical red nova like transient to a contact binary progenitor, that of V1309 Sco \citep{2011A&A...528A.114T}. Other examples such as V4432 Sgr \citep{1999AJ....118.1034M}, V838 Mon \citep{2002IAUC.7785....1B} and OGLE2002-BLG-360 \citep{2013A&A...555A..16T} are postulated to represent stellar mergers although their progenitors remain unidentified. In addition there exist possible historical and extra galactic examples \citep{2019A&A...630A..75P, 2014MNRAS.443.1319K}. Since the recognition of V1309 Sco there has been heightened interest in the theoretical basis of orbital instability and the identification of low mass ratio contact binary systems \citep{2021MNRAS.501..229W,2021MNRAS.502.2879G,2022MNRAS.tmp..527C}.

In two papers \citet{1993PASP..105.1433R,2001AJ....122.1007R} showed that the shape of contact binary light curves is dependant on three main geometrical factors namely the mass ratio ($q$), the fill-out ($f$) (degree of contact - or the thickness of the contact neck region) and the inclination ($i$). In addition the other somewhat minor determinant is the temperature difference between the components. He deduced that the maximum amplitude for any given system was seen if the system was observed to have a complete eclipse. In addition, he showed that only the mass ratio, fill-out and to a lesser extent (in the presence of complete eclipses) inclination determined the amplitude of the light curve with other factors such as temperature of the components having little impact.

We combine the three techniques noted above namely the theoretical instability parameters, survey photometric data and the photometric amplitude distribution of contact binary systems to derive a simplified method of identifying potential contact binary systems that show signs of orbital instability (potential red nova progenitors). For the purpose of this study we define a contact binary system with a mass ratio at or below the theoretical critical level as being potentially unstable. The paper is divided into five sections. In section two we model theoretical light curves to derive a relationship between the mass of the primary and the maximum amplitude of a potentially unstable system. In section 3 we employ the relationship to the ASAS-SN survey to identify potential bright red nova candidates among low mass ($0.6M_{\odot} < M_1 < 1.4M_{\odot}$) contact binary systems. %In section 4 we perform light curve analysis of the identified targets. %We  reviewed all the data sources identified in The International Variable Star Index (VSX) \citep{2006SASS...25...47W} along with The Transiting Exoplanet Survey Satellite (TESS) \citep{2021ApJS..254...39G} to select the sources with maximum phase coverage and low scatter. In most cases where available the data from TESS mission was used. As an added confirmation of the suitability of survey data we acquired ground based multi-band light curves on a sub-sample and compared light curve solutions relative to survey photometry solutions. 
In section 4 we define an average potential red nova progenitor in addition to comparing and contrasting our sample of bright potential red nova progenitors with other comparable systems. In section 5 we briefly discuss the historical development of mass ratio as a determinant of orbital instability, limitations of the present study and further ongoing search of red nova progenitors along with a summary and conclusion of the current work. 

\section{Amplitude Distribution of Potential Red Nova Progenitors}
Critical to understanding the orbital evolution of contact binary system is predicated on knowledge of parameters such as the mass ratio, masses of the components and the geometry of the orbit  such as inclination, degree of contact and temperature variation between the components. Many, if not all, of these parameters can be derived from light curve analysis but only if the light curve demonstrates a complete eclipse \citep{2005Ap&SS.296..221T}. As such, we limit our modelling of systems that demonstrate a complete eclipse and would be applicable to observed light curves

Current light curve analysis tools can incorporate many tens of different parameters, however, as noted above in the case of contact binary systems only 4 main parameters are critical. Therefore in modelling the amplitude distribution of unstable systems we have neglected complications associated with star spots and other stellar activity. We used the 2009 version of the Wilson-Devinney code as incorporated into the Windows front end utility WDwin56d \citep{2021NewA...8601565N} to model all light curves. The gravity darkening coefficients $g_1 = g_2 = 0.32$, the bolometric albedos $A_1 = A_2 = 0.5$ were fixed \citep{1967ZA.....65...89L} and simple reflection treatment applied \citep{1969PoAst..17..163R}. As per \citet{2015IBVS.6134....1N} logarithmic limb darkening coefficients interpolated from \citet{1993AJ....106.2096V} were used. 

To confirm the findings of \citet{1993PASP..105.1433R,2001AJ....122.1007R} with respect to the effects of fill-out and inclination on the amplitude of the light curve we modelled light curves of an idealised contact binary system with primary of one solar mass with mass ratio ($q$) 0.1 and equal temperature of the components ($T_1 = T_2 = 5770K$) to record the effects of inclination ($i$) and degree of contact ($f=0 - 1$). As we are only interested in systems that display a total eclipse we modelled the system with an inclinations of $90^{\circ}$ which would yield the maximum eclipse time and $72^{\circ}$ which would yield a small total eclipse between phase 0.49 to 0.51. The change in inclination results in a reduction in the duration of the secondary eclipse with the lower inclination reducing the total amplitude slightly. We also modelled fill-out $f=0$ and $f=1$ for each inclination to determine the variation in amplitude and again to confirm previous findings using the current accepted modelling code. The fill-out has a significant effect on the amplitude of the light curve with high fill-out yielding the highest amplitude. This is as expected, because  higher the degree of contact the thicker the neck of the contact region. The neck bears some luminosity and thicker the neck the greater the luminosity that is eclipsed. We did not model stars of different mass (hence different $T_1$) because as noted by \citet{2001AJ....122.1007R} the combination chosen is a reasonable representation of the light curve of low mass contact binaries and confirms the previous findings that the maximum amplitude of a contact binary system occurs at high inclination and high degree of contact. The results are illustrated in Figure 1 and summarised in Table 1.

Having established the condition of high inclination and high fill-out for high amplitude we next modelled the effects of the difference in temperature of the components. The presence of a common envelope usually results in good thermal contact between the components so there is usually little difference in the temperatures of the components. Recently \citet{2021ApJS..254...10L} compiled a catalogue of published light curve solutions of contact binaries. Using a sub-sample of the catalogue for primary stars between ($0.6M_{\odot} < M_1 < 1.4M_{\odot}$) we determined the median temperature difference between the components of approximately $200K$. Adopting the commonest temperature difference between the components we next modelled light curves for our idealised system with ($i=90^{\circ}$ and $f=1$) and temperatures of the secondary either 200K higher ($5970K$) or 200K lower ($5570K$). The results are illustrated in Figure 1 and summarised in Table 1. It is clear that the typical temperature difference has little effect on the maximum amplitude of the light curve with both the cooler or warmer secondaries having minimal impact relative to the result with the components being of same temperature. We can deduce from the above that for any given mass ratio the maximum amplitude will be achieved with high inclination, high fill-out and the secondary slightly warmer.

\begin{table}[ht]
    \centering
    \begin{tabular}{|c|c|c|c|}
    \hline
        Inclination($^\circ$) &fill-out&$T_2$& Max Ampl (Mag) \\ \hline
        90&1& Eq  & 0.36\\ \hline
        72&1& Eq & 0.31\\ \hline
        90&0& Eq & 0.27\\ \hline
        90&1&CS & 0.36\\ \hline
        90&1& HS & 0.36\\ \hline
    \end{tabular}
    	\caption{Effects of inclination, fillout and temperature of the secondary on the maximum amplitude of a contact binary system with mass ratio 0.1. Eq = Equal component temperatures, CS = Cold Secondary, HS = Hot secondary, $r_{1,2}$ = mean fractional radii of primary and secondary.}
\end{table}

\begin{figure*}[ht]
    \label{fig:JAAFIG1}
	\includegraphics[width=\textwidth]{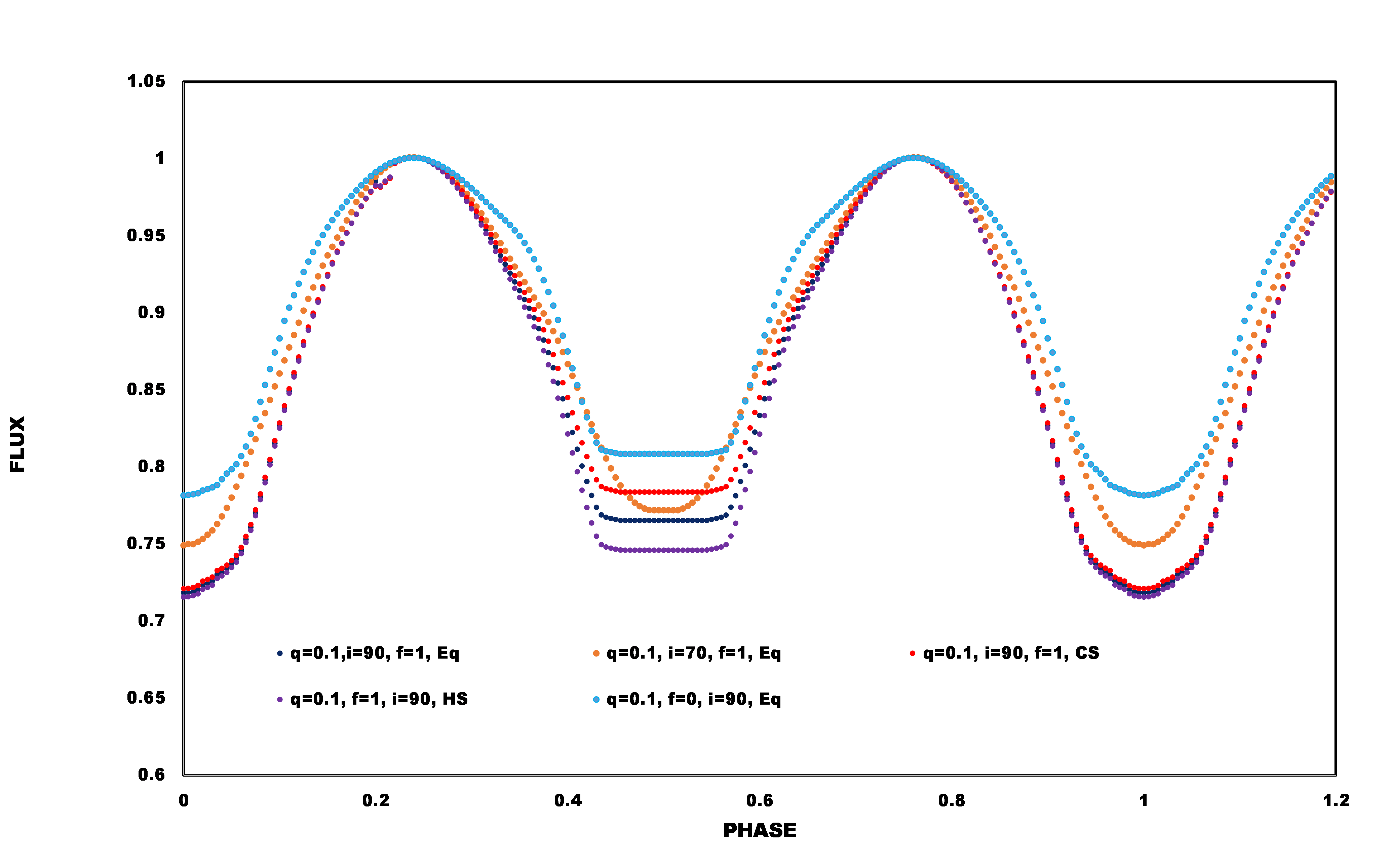}
    \caption{Effects of inclination, fillout and temperature of the secondary on the maximum amplitude of a contact binary system with mass ratio 0.1. Eq = Equal component temperatures, CS = cooler secondary, HS = hotter secondary. }
    \end{figure*}

Having establish the modelling criteria for the maximum amplitude at a given mass ratio we employ this to determine the maximum amplitude of a potential red nova progenitor. Recently \citet{2021MNRAS.501..229W} linked the instability mass ratio of contact binary systems with the mass of the primary component. They demonstrated that the mass ratio at which instability ($q_{inst}$) is likely can be determined by a simple quadratic relationship for high and low level of contact:
\begin{equation}
\label{eq:qinst-f1}
    q_{inst}=0.1269M_{1}^2-0.4496M_{1}+0.4403\ (f=1). 
    \end{equation}
    \begin{equation}
\label{eq:qinst-f0}
  q_{inst}=0.0772M_{1}^2-0.3003M_{1}+0.3237\ (f=0). 
    \end{equation}

Using this relationship say for a system with primary of one solar mass the instability mass ratio at high fill-out would be approximately 0.12. If we now model a light curve with the following parameters $T_1 = 5770K, T_2 = 5970K, i = 90^{\circ}, f = 1$ and $q = 0.12$ this will give us the maximum amplitude at which such a system is possibly unstable. As amplitude increases with increasing mass ratio (see below and \citet{1993PASP..105.1433R,2001AJ....122.1007R}), any system with amplitude higher will likely have a higher mass ratio and therefore would be stable. As inclination of a system drops eventually a complete eclipse is lost and light from both components is observed throughout the orbital cycle and the overall variation (amplitude) of the light curve drops \citep{2001AJ....122.1007R}. Therefore a system with amplitude significantly below the maximum amplitude is unlikely to have a complete eclipse and therefore not suitable for photometric light curve analysis. We extended our modelling of the maximum amplitude for systems with primary star masses between $0.6M_{\odot}$ and $1.4M_{\odot}$ and the calculated instability mass ratio at high fill-out. We adopted values of $T_1$ based on the main sequence calibration from \citep{2013ApJS..208....9P} + 200K. The results are summarised in Table 2 and the line of best fit is as shown in Eq 3 and graphically shown in Figure 2. It is clear that higher the mass ratio the greater the amplitude therefore any system with an amplitude higher than that predicated at the theoretical instability mass ratio will likely have a mass ratio above the instability value and therefore be likely stable. 

\begin{table}[ht]
    \centering
    \begin{tabular}{|c|c|c|}
    \hline
        Mass ($M_1) (M_{\odot}$) &$q_{inst} (f=1)$&Max Ampl\\ \hline
        0.6&0.22&0.63\\ \hline
        0.7&0.19&0.60\\ \hline
        0.8&0.16&0.52\\ \hline
        0.9&0.14&0.45\\ \hline
        1.0&0.12&0.43\\ \hline
        1.2&0.08&0.32\\ \hline
        1.4&0.06&0.22\\ \hline
    \end{tabular}
    	\caption{Summary of the maximum amplitude at the instability mass ratio ($q_{inst}$) for systems with low mass primary component.}
\end{table}
 
\begin{equation}
\label{eq:maxAmpl}
   MaxAmpl\ (mag)=-0.5179M_1 + 0.945 
\end{equation}

\begin{figure}[ht]
    \label{fig:JAAFIG2}
	\includegraphics[width=\columnwidth]{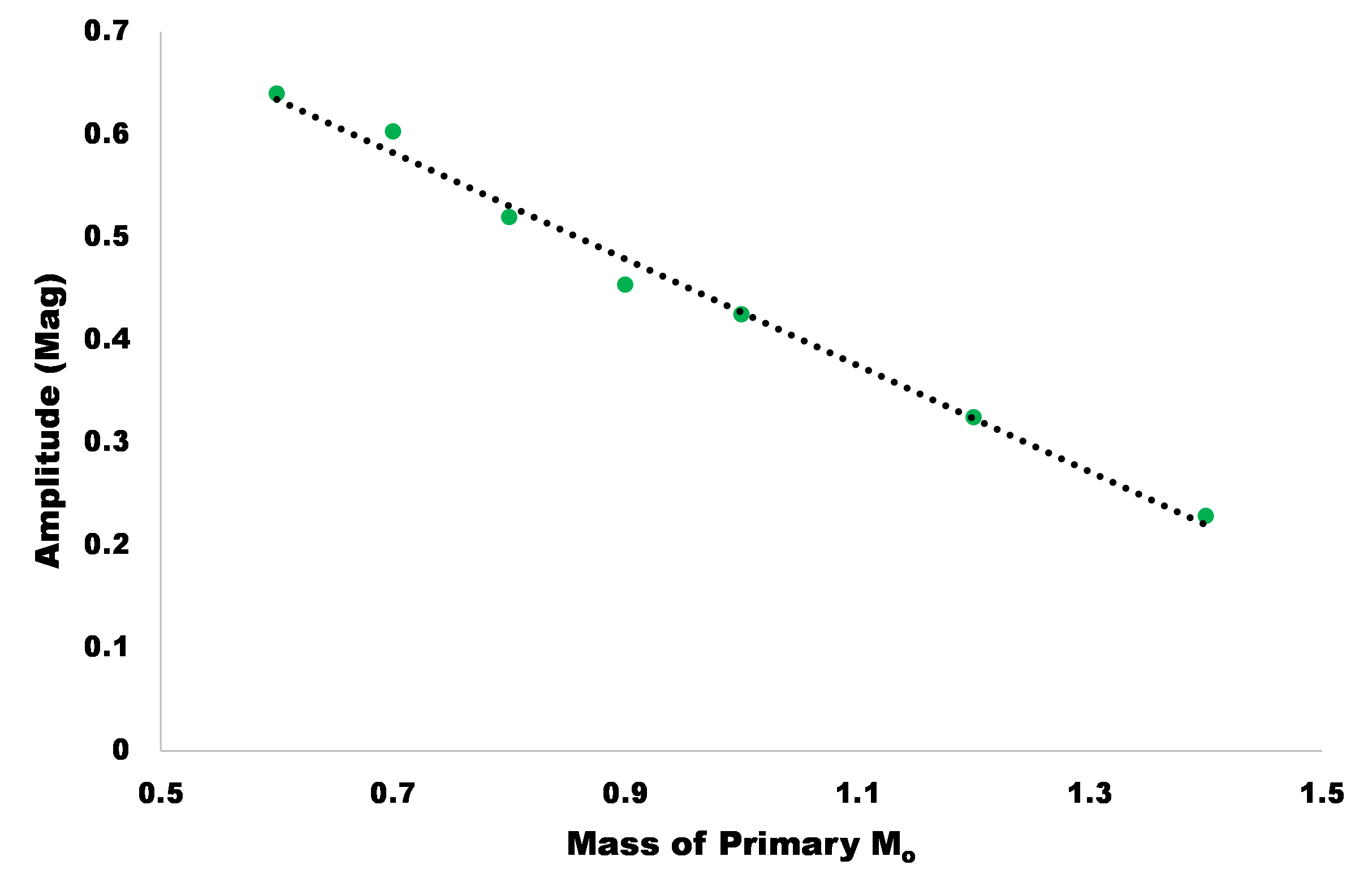}
    \caption{Maximum amplitude of contact binary systems with mass of the primary between $0.6M_{\odot}$ and $1.4M_{\odot}$ and mass ratio at the instability level ($f$=1).}
    \end{figure}

\section{Search for Potential Red Nova Progenitors}
As noted above we confined our search to bright systems with lower mass ($0.6M_{\odot} < M_1 < 1.4M_{\odot}$) primaries. The ASAS-SN variable database provides a friendly user interface to select variable stars of different types. We used this to select all contact binaries brighter than 13.5 magnitude. We limited our search to brighter examples as random review of the ASAS-SN light curves indicated that fainter examples had too much scatter and in most cases it was impossible to visually confirm the presence of complete eclipses and/or the amplitude of the light curve. A secondary benefit for favouring brighter examples is the potential ease in obtaining long term follow up monitoring. The selected systems are within the reach of modest instruments and could potentially be observed regularly by campus based telescopes or even by advanced amateurs. The systems were then ordered by amplitude and all systems with amplitude greater than 0.65 were excluded as the maximum amplitude for a 0.6$M_{\odot}$ primary with a mass ratio at the instability level ($q_{inst} = 0.22)$ is 0.63 magnitudes. 

It is well established that the primary component of a contact binary systems follows in general a main sequence profile \citep{2013MNRAS.430.2029Y}. The ASAS-SN database also provided the J and H magnitude for each system. We calculated the J-H magnitude for each system and using the calibration for low mass (F3-K9) main sequence stars of \citet{2013ApJS..208....9P} we interpolated the mass and effective temperature of the primaries. All systems with masses outside our inclusion criteria were excluded. The remaining systems were examined visually to determine the presence of a total eclipse. Only those systems where a clear total eclipse were included. We note that such a crude selection system is likely to include or exclude some systems in the final sample. A more robust system of checking the variability within a defined time interval of the eclipses proved unworkable due to the cadence and scatter in many systems. The final sample totaled 189 contact binary systems. Having established the mass of the primaries for our selected 189 samples we next determined the instability mass ratio for each using Eq 1 and the maximum amplitude using Eq 3. We next compared the amplitude of the survey light curve against the maximum amplitude for potential instability. If the observed amplitude was significantly ($>5\%$) higher than the maximum instability amplitude than such a system would be expected to have a mass ratio higher than the instability mass ratio and not be considered a potential merger candidate. All systems identified as such were excluded. The 5\% leeway is arbitrary to account for the scatter normally present in survey data. This step left a sample of 65 systems of potential red nova progenitors. 

Even though the ASAS-SN light curves were of reasonable quality we searched the VSX database as well as the TESS and Kepler variable databases for all available light curves for the 65 systems. All light curves found from other databases were compared with the available ASAS-SN light curves and if they offered better phase coverage and clearer eclipses they were chosen for the formal light curve analysis instead of the ASAS-SN curves. Although where TESS data was available it provided the cleanest curves we did in some cases use the ASAS-SN, Catalina survey \citep{2017MNRAS.469.3688D} and SWASP Survey \citep{2006PASP..118.1407P} photometry.

All the selected light curves from the final sample were analysed using the Wilson-Devenney code as noted above. We used the standard mass ratio search grid method to find the probable mass ratio for each system. Temperature of the primary was fixed according to the main sequence calibration of \citet{2013ApJS..208....9P}.
%Temperature of the primary was fixed according to the main sequence calibration of \citet{2013ApJS..208....9P}. The temperature of the primary has no specific impact on the light curve solution as the light curve shape is govern almost exclusively by geometric parameters \citep{1993PASP..105.1433R}. 
Logarithmic limb darkening coefficients were interpolated from \citet{1993AJ....106.2096V}. The TESS and Kepler photometric data is provided as a flux and this was converted to magnitudes using the calibrations from \citet{2021AJ....162..170H} and \citet{2015MNRAS.447.2880A} respectively. We used the MIT Quick Look pipeline for the optimal aperture for the TESS data \citep{2020RNAAS...4..204H} and K-2 data when using Kepler mission data. The TESS photometry was acquired over a broad red to infra-red window centered on the standard $I_c$ band (786.5nm) \citep{2015JATIS...1a4003R}. We used limb darkening coefficients for the $I_c$ band when analysing the TESS data. The Kepler and SWASP photometry was acquired with wide-band filters from blue to red \citep{2010AAS...21542002V, 2006PASP..118.1407P} and we used limb darkening coefficients for the central V band.

Given the scatter and potential incomplete phase coverage some of the systems upon analysis had a mass ratio above the instability mass ratio suggesting that the amplitude is probably higher than that measured by survey photometry. As noted by \citet{2022MNRAS.tmp..527C} a small uncertainty in the mass of the primary can result in a modest uncertainty in the instability mass ratio. Accordingly, for the purpose of this study we consider any system with a mass ratio below or up to 10\% above the theoretical maximum instability mass ratio to be potentially a red nova candidate. From our initial sample of 65 there were 45 that met the instability criteria and the basic parameters of these are summarised in Table 3. Abbreviation and cross matching of individual systems are presented in Table 5.  The list will of course in the future be refined as the mass of the primary of these systems is more accurately determined. It must be stressed that this study only covers the ASAS-SN variable database. Given the sheer number of bright contact binaries (magnitude $\leq 13.5$) listed on the VSX ($\approx 20000$) at this time we have not systematically reviewed other survey data (some with poor search interfaces) for more examples. We hope to do this over time and add to the list progressively. For completeness we do add some examples to the list from the existing literature as described below.

\section{Absolute Parameters and the Average Potential Red Nova Progenitor}
\subsection{Absolute Parameters}

We determined the absolute parameters for each system from the light curve solution and period of the system. As noted above mass of the primary was estimated from the J-H colour of each system. The mass ratio provides the mass of the secondary. The relative radii of the components are dependant on the mass ratio and Roche geometry as by definition both components overflow their inner Roche lobes in a contact binary system. The light curve solution provides fractional radii of the components ($a_{1,2},b_{1,2},c_{1,2}$) in three orientations. The geometric mean of these was used to estimate $r_{1,2}= \sqrt[3]{a_{1,2}b_{1,2}c_{1,2}}$. The separation ($A$) between the components was determined using Kepler's third law and the absolute radii of the components were determined as per \citep{2005JKAS...38...43A} $R_{1,2}$ = $A\times r_{1,2}$.

By way of comparison with other contact binary systems we looked at our list of possible red nova progenitors with low mass ($0.6M_{\odot}<M_1<1.4M_{\odot}$) primary contact binary systems listed by \citet{2021ApJS..254...10L}.  We accepted as true the masses determined by the publishing authors regardless of the methodology employed. Allowing for a 10\% margin the final list of 300 systems includes 9 systems that would be classified as potential red novas progenitors based on the instability criteria outlined above. Of these 9, three were already included in our list while the others were either fainter than our cut-off limit, had poor phase coverage and in two cases were too bright for the ASAS-SN survey equipment. We have added these to our final list of potential red nova progenitors relying on the published absolute parameters. As the catalogue of \citet{2021ApJS..254...10L} covers literature to the early part of 2021 we performed a literature search from March 2021 to March 2022 for any new reported contact binary systems that may be
regarded as red nova progenitors. This resulted in the addition three more potential systems. In total we catalogue 54 low mass ratio contact binary systems that maybe regarded as potential red nova progenitors (Table 3).

 \begin{table*}
    \centering
    \scriptsize
    \begin{center}

    \begin{tabular}{|l|l|l|l|l|l|l|l|l|l|l|}
    \hline
        Name & Period & $q$ & $q_{inst}$ range & $T_1$ & $T_2$ & $M_1(M_{\odot})$ & $R_1 (R_{\odot})$ & $R_1/ZAMS$ & Survey & References\\ \hline
        A0006 & 0.38318 & 0.115 & 0.108 - 0.128 & 5700 & 5699 & 0.95 & 1.34 & 1.39 & TESS & ~\\ \hline
        LM Psc & 0.34013 & 0.096 & 0.082 - 0.094 & 6075 & 6241 & 1.13 & 1.33 & 1.19 & ASAS-SN & ~\\ \hline
        A0346 & 0.30717 & 0.148 & 0.139 - 0.171 & 5120 & 5043 & 0.77 & 1.05 & 1.29 & TESS & ~\\ \hline
        A0458 & 0.33348 & 0.086 & 0.082 - 0.093 & 6100 & 5625 & 1.14 & 1.3 & 1.16 & TESS & ~\\ \hline
        A0514 & 0.34572 & 0.127 & 0.116 - 0.138 & 5600 & 5626 & 0.9 & 1.22 & 1.32 & TESS & ~\\ \hline
        NSVS 470 & 0.35576 & 0.078 & 0.095 - 0.110 & 5900 & 6231 & 1.04 & 1.37 & 1.32 & TESS & ~\\ \hline
        V644 Pup & 0.33056 & 0.14 & 0.132 - 0.161 & 5300 & 5824 & 0.8 & 1.12 & 1.33 & TESS & ~\\ \hline
        A0842 & 0.33353 & 0.1 & 0.086 - 0.098 & 6040 & 6315 & 1.1 & 1.3 & 1.23 & TESS & ~\\ \hline
        A1037 & 0.34370 & 0.09 & 0.070 - 0.085 & 6200 & 6081 & 1.21 & 1.34 & 1.14 & CATALINA & ~\\ \hline
        A1214 & 0.39850 & 0.085 & 0.099 - 0.116 & 5850 & 5786 & 1.01 & 1.43 & 1.41 & KEPLER & ~\\ \hline
        A1249 & 0.37191 & 0.095 & 0.091 - 0.104 & 5950 & 5948 & 1.07 & 1.39 & 1.3 & TESS& ~\\ \hline
        A1251 & 1.05207 & 0.085 & 0.073 - 0.082 & 6200 & 5599 & 1.21 & 2.89 & 2.45 & TESS & ~\\ \hline
        SSS1315 & 0.38281 & 0.075 & 0.108 - 0.128 & 5700 & 5330 & 0.95 & 1.39 & 1.44 & CATALINA & ~\\ \hline
        A1407 & 0.36358 & 0.088 & 0.086 - 0.098 & 6040 & 6105 & 1.1 & 1.39 & 1.27 & TESS & ~\\ \hline
        A1446 & 0.35170 & 0.09 & 0.116 - 0.138 & 5600 & 5369 & 0.9 & 1.28 & 1.39 & TESS & ~\\ \hline
        A1517 & 0.32518 & 0.1 & 0.132 - 0.161 & 5300 & 5368 & 0.8 & 1.13 & 1.34 & ASAS-SN & ~\\ \hline
        A1531 & 0.83309 & 0.085 & 0.082 - 0.093 & 6100 & 5812 & 1.14 & 2.43 & 2.17 & KEPLER & ~\\ \hline
        V396 Lup & 0.36324 & 0.132 & 0.120 - 0.145 & 5500 & 5794 & 0.88 & 1.27 & 1.39 & TESS & ~\\ \hline
        A1629 & 0.31077 & 0.059 & 0.104 - 0.122 & 5800 & 6115 & 0.98 & 1.22 & 1.23 & ASAS & ~\\ \hline
        A1651 & 0.35321 & 0.152 & 0.129 - 0.157 & 5300 & 5158 & 0.82 & 1.19 & 1.38 & TESS & ~\\ \hline
        V565Dra & 0.39032 & 0.092 & 0.091 - 0.104 & 5970 & 6055 & 1.07 & 1.42 & 1.34 & TESS & ~\\ \hline
        A1751 & 0.93521 & 0.105 & 0.095 - 0.110 & 5900 & 5551 & 1.04 & 2.46 & 1.49 & ASAS & ~\\ \hline
        A1846 & 0.30284 & 0.162 & 0.130 - 0.160 & 5250 & 5148 & 0.8 & 1.06 & 1.25 & KEPLER & ~\\ \hline
        A1847 & 0.68961 & 0.077 & 0.077 - 0.087 & 6150 & 5917 & 1.18 & 2.18 & 1.89 & ASAS-SN & ~\\ \hline
        A1907 & 0.76308 & 0.06 & 0.093 - 0.106 & 5970 & 5619 & 1.07 & 2.32 & 2.18 & ASAS-SN & ~\\ \hline
        A1928 & 0.32002 & 0.08 & 0.082 - 0.093 & 6100 & 6120 & 1.14 & 1.26 & 1.13 & SWASP & ~\\ \hline
        A1946 & 0.38378 & 0.1 & 0.120 - 0.144 & 5520 & 5524 & 0.88 & 1.3 & 1.43 & ASAS-SN & ~\\ \hline
        A2001 & 0.35088 & 0.12 & 0.123 - 0.140 & 5450 & 5614 & 0.86 & 1.18 & 1.33 & ASAS-SN & ~\\ \hline
        A2003 & 0.45720 & 0.127 & 0.107 - 0.128 & 5720 & 5613 & 0.95 & 1.52 & 1.58 & ASAS & ~\\ \hline
        A2044 & 0.34289 & 0.103 & 0.099 - 0.116 & 5800 & 5751 & 1 & 1.29 & 1.28 & TESS & ~\\ \hline
        A2044 & 0.37054 & 0.07 & 0.082 - 0.093 & 6100 & 6076 & 1.14 & 1.45 & 1.29 & ASAS-SN & ~\\ \hline
        NSVS 114 & 0.35470 & 0.096 & 0.116 - 0.138 & 5580 & 5572 & 0.9 & 1.28 & 1.38 & ASAS-SN & ~\\ \hline
        A2048 & 0.80583 & 0.115 & 0.104 - 0.122 & 5800 & 5870 & 0.98 & 2.2 & 2.23 & SWASP & ~\\ \hline
        A2132 & 0.31634 & 0.1 & 0.108 - 0.127 & 5700 & 5953 & 0.95 & 1.18 & 1.22 & TESS & ~\\ \hline
        A2145 & 0.35385 & 0.075 & 0.099 - 0.116 & 5850 & 5770 & 1.01 & 1.32 & 1.3 & TESS & ~\\ \hline
        SSS 2213 & 0.36990 & 0.08 & 0.082 - 0.093 & 6100 & 6249 & 1.14 & 1.42 & 1.26 & TESS & ~\\ \hline
        A2222 & 0.40609 & 0.133 & 0.116 - 0.138 & 5580 & 5853 & 0.9 & 1.38 & 1.49 & TESS & ~\\ \hline
        A2243 & 0.33535 & 0.085 & 0.091 - 0.104 & 6000 & 6004 & 1.07 & 1.29 & 1.22 & TESS & ~\\ \hline
        A2250 & 0.31196 & 0.132 & 0.126 - 0.153 & 5400 & 5584 & 0.84 & 1.11 & 1.26 & SWASP & ~\\ \hline
        A2258-26 & 0.32764 & 0.093 & 0.084 - 0.095 & 6050 & 5955 & 1.12 & 1.29 & 1.17 & SWASP & ~\\ \hline
        A2258 & 0.30946 & 0.162 & 0.129 - 0.157 & 5300 & 5496 & 0.82 & 1.05 & 1.22 & ASAS-SN & ~\\ \hline
        NSVS 902 & 0.32510 & 0.108 & 0.104 - 0.122 & 5800 & 5718 & 0.98 & 1.19 & 1.21 & SWASP & ~\\ \hline
        A2348 & 0.34719 & 0.108 & 0.091 - 0.104 & 5970 & 5807 & 1.07 & 1.29 & 1.22 & SWASP & ~\\ \hline
        V1222 Tau & 0.29536 & 0.104 & 0.116 - 0.138 & 5600 & 5439 & 0.9 & 1.06 & 1.15 & LIT & \citet{2015PASJ...67...74L}\\ \hline
        NSVS 431 & 0.25596 & 0.147 & 0.151 - 0.188 & 6000 & 5875 & 0.7 & 0.87 & 1.14 & LIT & \citet{2020NewA...7701352K}\\ \hline
        A0822 & 0.28005 & 0.11 & 0.087 - 0.099 & 5960 & 6080 & 1.1 & 1.1 & 1.01 & LIT & \citet{2015MNRAS.446..510K}\\ \hline
        GSC 0341 & 0.27716 & 0.055 & 0.092 - 0.106 & 5870 & 5828 & 1.06 & 1.17 & 1.1 & LIT & \citet{2021ApJ...922..122L}\\ \hline
        A0832 & 0.31132 & 0.067 & 0.072 - 0.081 & 6300 & 6602 & 1.22 & 1.34 & 1.13 & LIT & \citet{2016AJ....151...69S}\\ \hline
        NSVS 780 & 0.28120 & 0.098 & 0.141 - 0.173 & 5490 & 5706 & 0.79 & 1.1 & 1.31 & LIT & \citet{2021RAA....21..225P}\\ \hline
        PZ UMA & 0.26267 & 0.178 & 0.139 - 0.170 & 5430 & 4972 & 0.77 & 0.92 & 1.12 & LIT & \citet{2019PASJ...71...39Z}\\ \hline
        NSVS 256 & 0.28780 & 0.078 & 0.078 - 0.088  & 6030 & 6100 & 1.17 & 1.19 & 1.04 & LIT & \citet{2018RAA....18..129K}\\ \hline
        SX Crv & 0.31662 & 0.079 & 0.069 - 0.077 & 6340 & 6160 & 1.25 & 1.32 & 1.09 & LIT & \citet{2004AcA....54..299Z}\\ \hline
        A1328 & 0.38470 & 0.086 & 0.071 - 0.079 & 6300 & 6319 & 1.23 & 1.49 & 1.25 & LIT & \citet{2021ApJ...922..122L}\\ \hline
        ZZ PsA & 0.37389 & 0.078 & 0.086 - 0.098 & 6514 & 6703 & 1.213 & 1.42 & 1.24 & LIT & \citet{2021MNRAS.501..229W}\\ \hline
    \end{tabular}
    \caption{Summary of the pertinent light curve solution and absolute parameters of potential red nova progenitors. Entries marked with "LIT"have been taken from the literature. They were not identified from examination of the ASAS-SN light curves as described. $q_{inst}$ range is the instability mass ratio from $f=0-1$.}
    \end{center}
\end{table*}

%All other systems reported in this study are essentially new discoveries. We accept the reported light curve solution almost entirely as any potential change in the mass of the primary would have resulted in a possible change in the temperature assigned. As noted by \citet{1993PASP..105.1433R} $T_1$ will have no significant effect on the light curve solution with respect to the photometric mass ratio and other major parameters such as inclination and fill-out. The parameter that does change with any significant change in mass of the primary is the mean fractional radius and any subsequent absolute estimation of the radius. We recalculate the fractional radius for each system based on the J-H mass of the primary. As can be seen from Table 1 other aspects of the light curve solution have no significant impact on the mean fractional radius.

The period distribution of low mass contact binary systems ($0.6M_{\odot} < M_1 < 1.4M_{\odot}$) as adopted from \citet{2021ApJS..254...10L} as a whole relative to potential red nova progenitors is illustrated in Figure 3. The median period for the entire sample is in the order of 0.330 days, only marginally less than 0.346 days for potential red nova progenitors. Most of the systems in both groups have periods between 0.25 and 0.5 days although the peak in the distribution around 0.35 days is more pronounced in the potential red nova progenitor sample. There is a hint of possibly some systems with higher periods being more common in the potential red nova progenitors group. The finding is in line with \citet{2022arXiv220201187K} who found that extreme mass ratio systems tended to have a slightly increased frequency of longer periods. 

\begin{figure}[ht]
    \label{fig:JAAFIG3}
	\includegraphics[width=\columnwidth]{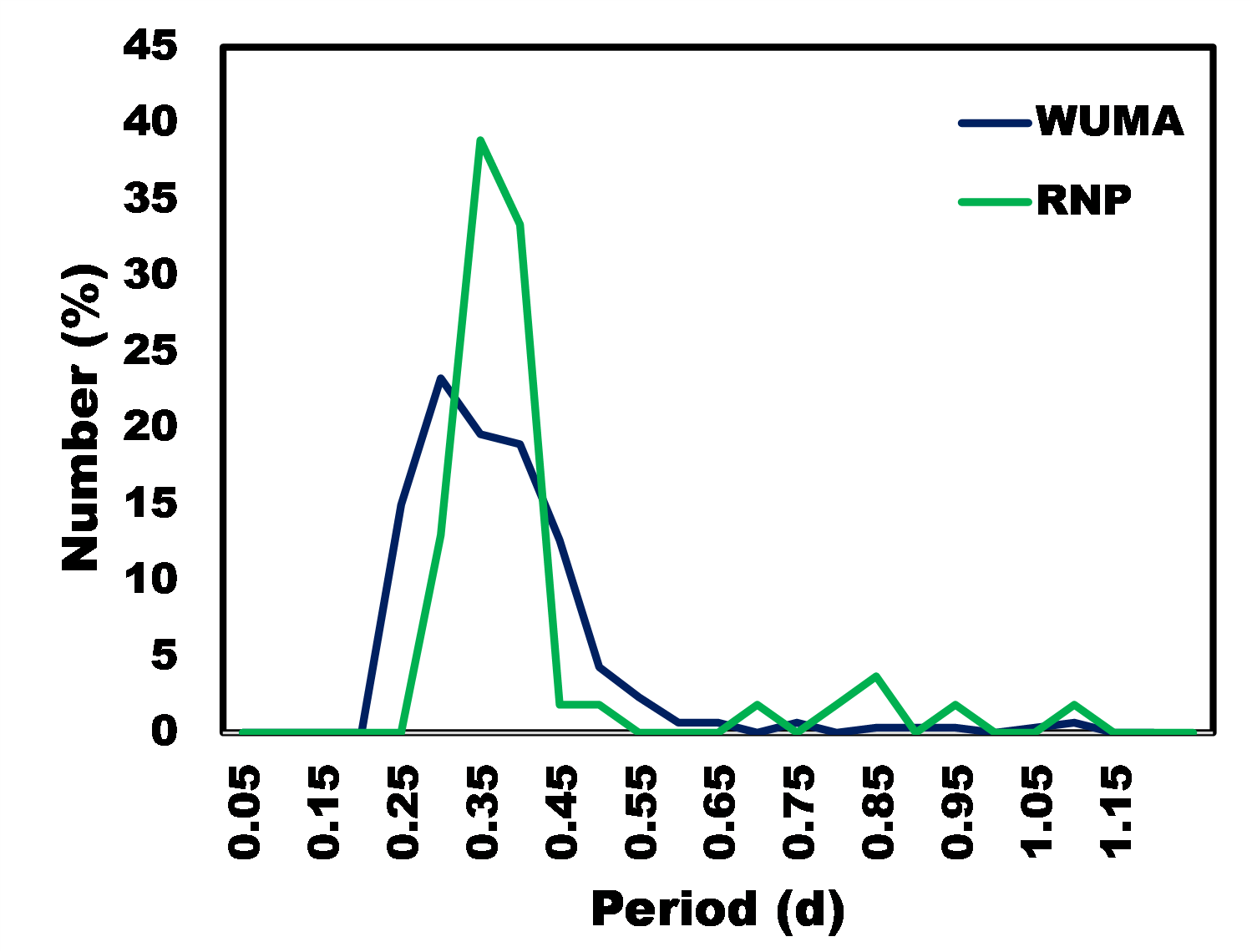}
    \caption{The period distribution of Red Nova Progenitors (RNP) relative to population of low mass contact binary systems (WUMA). Systems binned over 0.05 days}
    \end{figure}
\subsection{Average Potential Red Nova Progenitor}
Based on the work of \citet{2007MNRAS.377.1635A}, 
\citet{2021MNRAS.501..229W} the separation at the onset of instability ($A_{inst}$) can be written as:
%\begin{scriptsize}

\begin{eqnarray}
\label{eq:a-inst}
 \frac{A_{\mathrm{\scriptscriptstyle inst}}}{R_1} &=& \frac{q\frac{k_2^2}{k_1^2}{P}{Q}}{q \frac{k_2^2}{k_1^2}{P}^2 + \frac{q}{(1+q)k_1^2}}\\
 &+&\frac{\sqrt{(q\frac{k_2^2}{k_1^2}{P}{Q})^2 + 3 (1+q\frac{k_2^2}{k_1^2}{Q}^2) (q \frac{k_2^2}{k_1^2}{P}^2 + \frac{q}{(1+q)k_1^2})   }}{q \frac{k_2^2}{k_1^2}{P}^2 + \frac{q}{(1+q)k_1^2}}\nonumber
\end{eqnarray}
%\end{scriptsize}
\noindent where $k_{1,2}$ is the gyration radius for the primary and secondary components and
%\begin{footnotesize}

\begin{equation}
P = \frac{0.49q^{2/3}-3.26667q^{-2/3}(0.27q -0.12q^{4/3})}{0.6q^{2/3} + \ln (1+ q^{1/3})} ,
\end{equation}
%\end{footnotesize}

%\begin{footnotesize}
\begin{equation}
Q = \frac{(0.27q -0.12q^{4/3})({0.6q^{-2/3} + \ln (1+ q^{-1/3})})}{0.15 (0.6q^{2/3} + \ln (1+ q^{1/3}))}.
\end{equation}
%\end{footnotesize}
Assuming our sample of potential red nova progenitors as representative we find the median radius ($\pm SD$) of the primary to be 1.285 ($\pm 0.3$) times the corresponding ZAMS equivalent. As noted previously the primary of contact binary systems can be considered as ZAMS so we can estimate the typical radius and hence the instability separation for a typical potential red nova progenitor. We perform this for low mass contact binary systems with low and high degree of contact as described in \citet{2021MNRAS.501..229W} and adopt the mean. Using equations outlined above and mean instability separation we estimate the mean period at the onset of instability ($P_{inst}$) for low mass contact binary systems. The results are summarised in Table 4. From these we derive simple quadratic relations (Figures 4 and 5) linking the mass of the primary with the instability separation and period as follows:
\begin{table}[ht]
    \centering
    \begin{tabular}{|c|c|c|}
    \hline
        Mass ($M_1) (M_{\odot}$) &$A_{inst} (M_\odot)$&$P_{inst} (d)$\\ \hline
        0.6&1.571&0.269\\ \hline
        0.7&1.738&0.293\\ \hline
        0.8&1.899&0.316\\ \hline
        0.9&2.054&0.339\\ \hline
        1.0&2.202&0.359\\ \hline
        1.1&2.339&0.378\\ \hline
        1.2&2.459&0.393\\ \hline
        1.3&2.556&0.402\\ \hline
        1.4&2.620&0.403\\ \hline
    \end{tabular}
    	\caption{Separation and Period at the onset of instability for modelled potential red nova progenitors}
\end{table}

\begin{equation}
    A_{inst} = -0.6766M_1^2 + 2.6932M_1 + 0.1878
\end{equation}
\begin{figure}[h!]
    \label{fig:JAAFIG4}
	\includegraphics[width=\columnwidth]{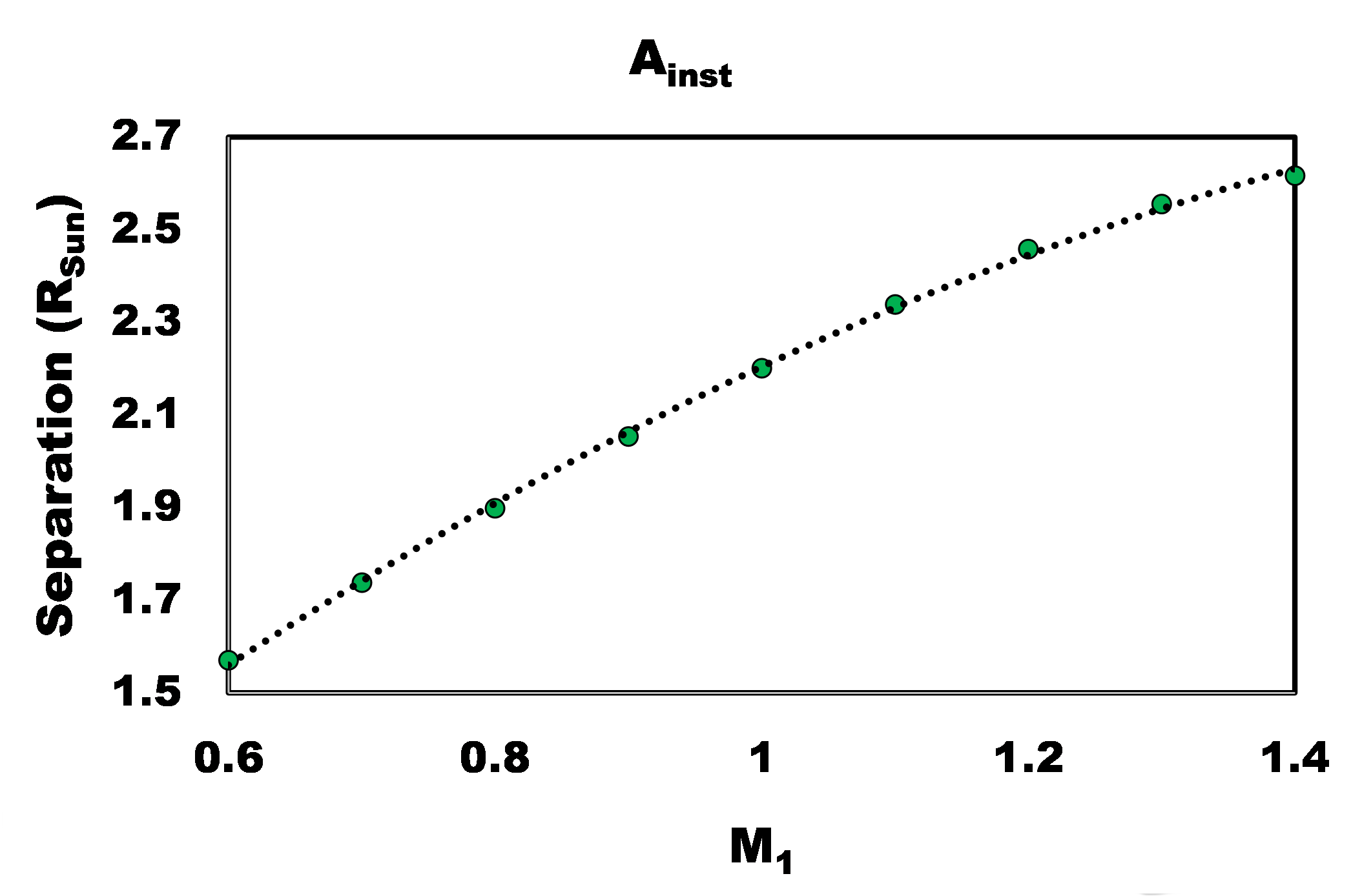}
    \caption{Instability separation for low mass contact binary systems.}
    \end{figure}
\begin{equation}
  P_{inst}(d) = -0.1446M_1^2 + 0.4645M_1 + 0.0401
\end{equation}
\begin{figure}[ht]
    \label{fig:JAAFIG5}
	\includegraphics[width=\columnwidth]{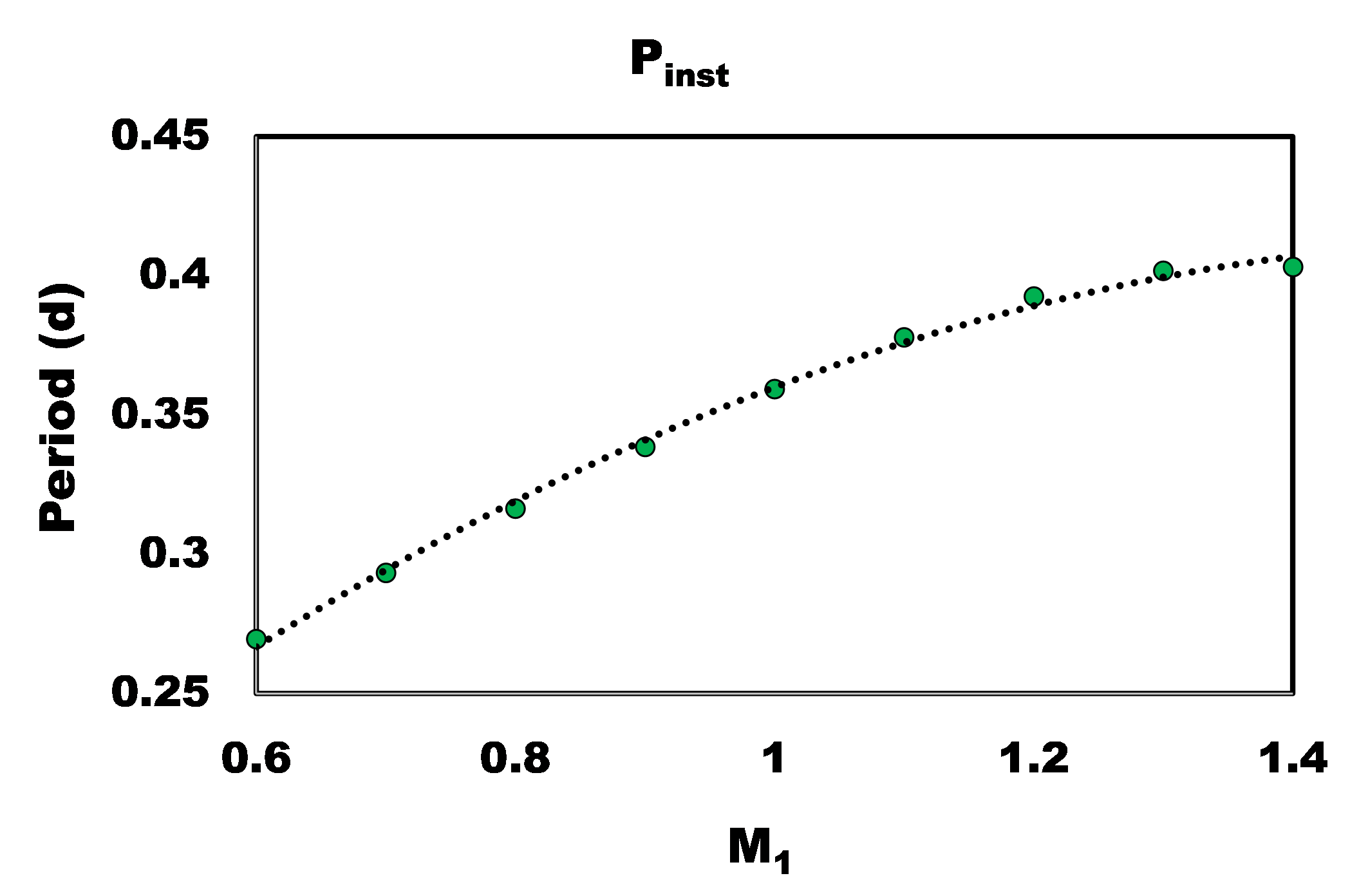}
    \caption{Instability period for low mass contact binary systems.}
    \end{figure}

From the above we see that a typical one solar mass contact binary system that is near the onset of instability will have a mass ratio near 0.12, maximum light curve amplitude near 0.43 and period near 0.36d. As a means of selecting potential low mass ratio contact binary stars for photometric analysis one could follow the relationships described to select candidates that are more likely to reflect features of orbital instability. Given the inherent scatter and varying cadence of survey photometry we stress the above analysis is only an aid in the selection process and follow up observations would be required to confirm instability criteria that maybe evident on survey photometry. We are however confident given the success of survey photometric analysis compared to dedicated ground based observations in determination of accurate light curve solutions \citep{2020ApJS..247...50S, 2020MNRAS.493.1565D, 2022arXiv220209120W} that if the survey data is chosen to ensure high cadence, low scatter and full phase coverage than the properties deduced from such data would be confirmed on follow-up study and hence are confident that the systems selected represent potential red nova progenitors.

\section{Discussion and Conclusions}
Although contact binary merger events are predicted to be relatively frequent so far only a single event has been confirmed and that only in retrospect. The linking of orbital stability with the mass ratio of contact binary systems has long been recognised as potential avenue for identifying unstable systems \citep{1995ApJ...444L..41R, 2007MNRAS.377.1635A, 2009MNRAS.394..501A}. The earlier work clearly showed that orbital instability is likely to occur at very low mass ratios and higher mass ratio configurations are likely to be stable. Our theoretical work \citep{2021MNRAS.501..229W} linking the instability mass ratio to the mass of the primary has progressed this further by demonstrating that there exists no global minimum mass ratio at which a system will become unstable rather the instability mass ratio is dependant on the mass of the primary component. In addition, we showed that systems with less massive primaries can have mass ratios higher than 0.2 and still be potentially unstable. When combined with work showing the suitability of survey photometry for light curve analysis \citep{2020ApJS..247...50S, 2020MNRAS.493.1565D, 2022arXiv220209120W} has greatly facilitated the potential for being able to to identify low mass ratio contact binary systems. In this study we enhance this capability by establishing simple light curve and colour parameters that can be used to exclude systems that are likely to have mass ratios above the theoretical instability value. By excluding likely stable systems we greatly increase our chance of identifying potentially unstable systems from the remaining sample. We apply the techniques on bright contact binary systems from the ASAS-SN and identify approximately 50 extreme low mass ratio system (most previously not reported) satisfying the mass ratio criteria for orbital instability. 

As with almost all low mass contact binary systems \citep{2021ApJS..254...10L} our identified sample of potential merger candidates demonstrate radii that are significantly larger than their main sequence counterparts and secondaries that are considerably hotter than main sequence counterparts of similar mass. We find that relative to the general population of comparable contact binaries those exhibiting signs of orbital instability generally have similar periods although with a more pronounced peak near 0.35d. In addition there is a significant number of systems, relative to the general population, that have longer ($>0.5d$) periods. \citet{2022arXiv220201187K} suggest that it is possible for the period to lengthen to above 0.5d in some cases of extreme low mass ratios and the onset of orbital instability. In this respect our sample of possible unstable systems with long periods represent a good subset study group for future observations.

From our sample of potential red nova progenitors we construct theoretical models for low mass potentially unstable contact binary systems which place further constraints on the light curve morphology and timings. We note that for low mass systems there is a narrow period domain from $\approx 0.27d$ to $0.4d$ at which they may become unstable and this increases with the mass of the primary. Those results will further aid in the identification of potential red nova progenitors from large survey samples. We hope to employ all the techniques and modifications described to the VSX database to further identify bright potential red nova candidates.

The techniques described in no way ensure all potential red nova progenitors are identified. Limiting our selection to those exhibiting complete eclipses clearly excludes a significant portion of contact binary systems, however, identification of potential systems among these would require both time consuming dedicated observations on modest sized telescopes and high resolution spectroscopic observations. Such requirements are unlikely to be readily available. The light curves of non totally eclipsing systems cannot be reliably analysed due to the high degree of correlation between the three geometric parameters specifically the inclination, mass ratio and degree of contact. The presence of a total eclipse places significant constraints, particularly in the $q/i$, domain allowing for manual search to be performed to find the correct light curve solution. Also, as already noted, survey photometric data although useful in light curve analysis does have limitations given the scatter, particularly with respect to the determining the  fill-out fraction \citep{2020MNRAS.493.1565D, 2022arXiv220209120W}. Looking at Equations 1 and 2 above we see that the fill-out can have a significant influence on the instability mass ratio particularly for systems with the primary below one solar mass. The methodology employed places sorting restrictions assuming a high fill-out so it is possible that some of the identified samples may still be in the stable range. The small sample identified offers a good opportunity for dedicated observations with modest optical instruments to further refine the list of potential merger candidates. To this end we have started a programme of dedicated multi-band observations of some of the identified systems with instruments in the 0.5m range with results to be presented progressively.

\begin{table*}

    \centering
    \begin{tabular}{|l|l|l|}
    \hline
        Abbreviation & RA/DEC & Name\\ \hline
        A0006 & 00 06 50 -35 37 29 & ASASSN-V J000649.98-353729.1\\ \hline
        LM Psc & 00 34 13 20 52 25 & ASASSN-V J003412.63+205225.4\\ \hline
        A0346 & 03 46 33.6 41 08 15.8 & ASASSN-V J034633.63+410815.8\\ \hline
        A0458 & 04 58 14 06 43 09 & ASASSN-V J045813.80+064309.1\\ \hline
        A0514 & 05 14 59 -73 56 15 & ASASSN-V J051459.38-735615.4\\ \hline
        NSVS 470 & 07 19 25 41 57 04 & ASASSN-V J071924.64+415705.4\\ \hline
        V644 Pup & 07 27 29 -50 56 30 & ASASSN-V J072728.92-505631.1\\ \hline
        A0842 & 08 42 20 -03 03 25 & ASASSN-V J084219.98-030325.3\\ \hline
        A1037 & 10 37 37 -37 09 30 & ASASSN-V J103736.72-370928.0\\ \hline
        A1214 & 12 14 30.5 -02 57 04 & ASASSN-V J121430.46-025704.6\\ \hline
        A1249 & 12 49 08 -29 44 38 & ASASSN-V J124907.83-294437.7\\ \hline
        A1251 & 12 51 19 -28 08 25 & ASASSN-V J125119.31-280824.8\\ \hline
        SSS1315 & 13 15 59.6 -37 00 17.7 & ASASSN-V J131559.62-370018.8\\ \hline
        A1407 & 14 07 13 -30 24 44 & ASASSN-V J140712.93-302443.8\\ \hline
        A1446 & 14 46 21 -30 04 40 & ASASSN-V J144620.72-300440.9\\ \hline
        A1517 & 15 17 02 14 10 23 & ASASSN-V J151701.56+141023.3\\ \hline
        A1531 & 15 31 18 -17 42 36 & ASASSN-V J153118.10-174236.0\\ \hline
        V396 Lup & 16 03 02 -37 49 21.2 & ASASSN-V J160302.12-374921.2\\ \hline
        A1629 & 16 29 19.9 35 40 03 & ASASSN-V J162919.96+354003.5\\ \hline
        A1651 & 16 51 39.4 22 55 44 & ASASSN-V J165139.40+225543.0\\ \hline
        V565Dra & 17 38 49.82 +57 12 23.2 & ASASSN-V J173849.79+571222.6\\ \hline
        A1751 & 17 51 10 03 13 20 & ASASSN-V J175109.86+031319.5\\ \hline
        A1846 & 18 46 43.4 -27 36 29 & ASASSN-V J184643.38-273629.3\\ \hline
        A1847 & 18 47 37 21 56 06 & ASASSN-V J184737.28+215606.0\\ \hline
        A1907 & 19 07 28.30 -53 47 24.7 & ASASSN-V J190728.21-534724.9\\ \hline
        A1928 & 19 28 49 -40 45 54 & ASASSN-V J192848.87-404554.0\\ \hline
        A1946 & 19 46 45 -04 03 39 & ASASSN-V J194644.82-040339.6\\ \hline
        A2001 & 20 01 26 07 37 40 & ASASSN-V J200125.92+073739.9\\ \hline
        A2003 & 20 03 04 -02 56 02 & ASASSN-V J200303.64-025603.3\\ \hline
        A2044 & 20 44 00 57 52 17 & ASASSN-V J204400.26+575216.7\\ \hline
        A2044 & 20 44 52 06 22 31 & ASASSN-V J204452.22+062231.3\\ \hline
        NSVS 114 & 20 45 26 16 59 13 & ASASSN-V J204525.65+165912.7\\ \hline
        A2048 & 20 48 35 -46 09 42 & ASASSN-V J204835.36-460942.4\\ \hline
        A2132 & 21 32 19.4 -53 51 33 & ASASSN-V J213219.30-535132.7\\ \hline
        A2145 & 21 45 37 -58 35 00 & ASASSN-V J214537.35-583459.9\\ \hline
        SSS 2213 & 22 13 27.3 -44 54 00.5 & ASASSN-V J221327.33-445400.3\\ \hline
        A2222 & 22 22 17 37 37 41 & ASASSN-V J222217.40+373740.6\\ \hline
        A2243 & 22 43 19 -73 51 18 & ASASSN-V J224318.80-735718.0\\ \hline
        A2250 & 22 50 00 -23 16 24 & ASASSN-V J224959.89-231623.1\\ \hline
        A2258-26 & 22 58 26 -26 03 36 & ASASSN-V J225825.91-260337.8\\ \hline
        A2258 & 22 58 50 13 49 18 & ASASSN-V J225849.67+134917.7\\ \hline
        NSVS 902 & 23 19 49 36 03 51 & ASASSN-V J231948.59+360350.6\\ \hline
        A2348 & 23 48 23 -40 54 41 & ASASSN-V J234823.30-405440.6\\ \hline
        V1222 Tau & 03 28 26 09 04 24 & V1222 Tau\\ \hline
        NSVS 431 & 04 59 45 49 25 03 & NSVS 4316778\\ \hline
        A0822 & 08 22 43 19 26 58 & ASASSN-V J082243.00+192658.5\\ \hline
        GSC 03415-02229 & 08 27 01 46 28 50 & GSC 03415-02229\\ \hline
        A0832 & 08 32 41 23 32 26 & ASASSN-V J083240.96+233225.9\\ \hline
        NSVS 780 & 09 06 43 70 03 29 & NSVS 780649\\ \hline
        PZ UMA & 09 29 07 49 51 23 & PZ UMA\\ \hline
        NSVS 256 & 10 10 42.7 67 39 31 & NSVS 2569022\\ \hline
        SX Crv & 12 40 15 -18 48 01 & SX Crv\\ \hline
        A1328 & 13 28 29 55 52 45 & ASASSN-V J132829.15+555245.4\\ \hline
        ZZ PsA & 21 50 35.2 -27 48 35.5 & ZZ PsA\\ \hline
            \end{tabular}
    \caption{Abbreviation, coordinates and ASAS-SN designation of low mass contact binary systems with mass ratios suggesting orbital instability.}
\end{table*}

\section*{Acknowledgements}

Based on data acquired on the Western Sydney University, Penrith Observatory Telescope. We acknowledge the traditional custodians of the land on which the Observatory stands, the Dharug people, and pay our respects to elders past and present.\\

BA acknowledges the financial support of the Ministry of Education, Science and Technological Development of the Republic of Serbia through the contract No.~451-03-68/2022-14/200104.\\

During work on this paper, G. Djurasevic and J. Petrovic were financially supported by the Ministry of Education and Science of the Republic of Serbia through contract 451-03-9/2021-14/200002.\\

This research has made use of the SIMBAD database, operated at CDS, Strasbourg, France.
\vspace{-1em}

%%use \balance somewhere in the left column of the last page to balance the two columns in the end page

%%References section
\bibliography{P1.bib}
%\begin{theunbibliography}{}
%\vspace{-1.5em}

%\bibitem{latexcompanion}
%Awadalla, N. S., & Hanna, M. A. 2005, Journal of Korean Astronomical Society, 38, 43
%\bibitem{latexcompanion}
%Dickey, J. M., Salpeter, E. E., Terzian, Y. 1978, Astrophys. J. Suppl. Ser., 36, 77
%\bibitem{latexcompanion}
%Radhakrishnan, G. C. {\em et al.} 1980, in Evans A., Bode M. F., eds, Non-Solar Gamma Rays (COSPAR), Pergamon Press, Oxford, p. 163
%\bibitem{latexcompanion}
%Starrfield S., Iliadis C., Hix W. R. 2008, in Bode M. F., Evans A., eds, Classical Novae, 2nd edition, Cambridge University Press, Cambridge, p. 77
%\bibitem{latexcompanion}
%Van Loon J. Th. 2008, in Evans A. et al., eds, R S Ophiuchi (2006) and the Recurrent Nova Phenomenon, ASP Conference Series, Volume 401, p. 90
%\bibitem{latexcompanion}
%Zwicky, F. 1957, Morphological Astronomy, Springer-Verlag, Berlin, p. 258

%\end{theunbibliography}

\end{document}